\documentclass[journal=nalefd,manuscript=letter]{achemso}

\usepackage{amsmath}
\usepackage{braket}
\usepackage{graphicx}% Include figure files
\usepackage{mathrsfs}
\usepackage{hyperref}
\usepackage{color}

\author{Huiyuan Zheng}
\altaffiliation{HKU-UCAS Joint Institute of Theoretical and Computational Physics at Hong Kong, Hong Kong, China}
\author{Dawei Zhai}
\altaffiliation{HKU-UCAS Joint Institute of Theoretical and Computational Physics at Hong Kong, Hong Kong, China}
\author{Wang Yao}
\affiliation[University of Hong Kong]{Department of Physics, The University of Hong Kong, Hong Kong, China}
\altaffiliation{HKU-UCAS Joint Institute of Theoretical and Computational Physics at Hong Kong, Hong Kong, China}
\email{wangyao@hku.hk}

\title
  {Anomalous magneto-optical response and chiral interface of dipolar excitons at twisted valleys}


\begin{document}

\begin{tocentry}
\includegraphics[width=0.8\textwidth]{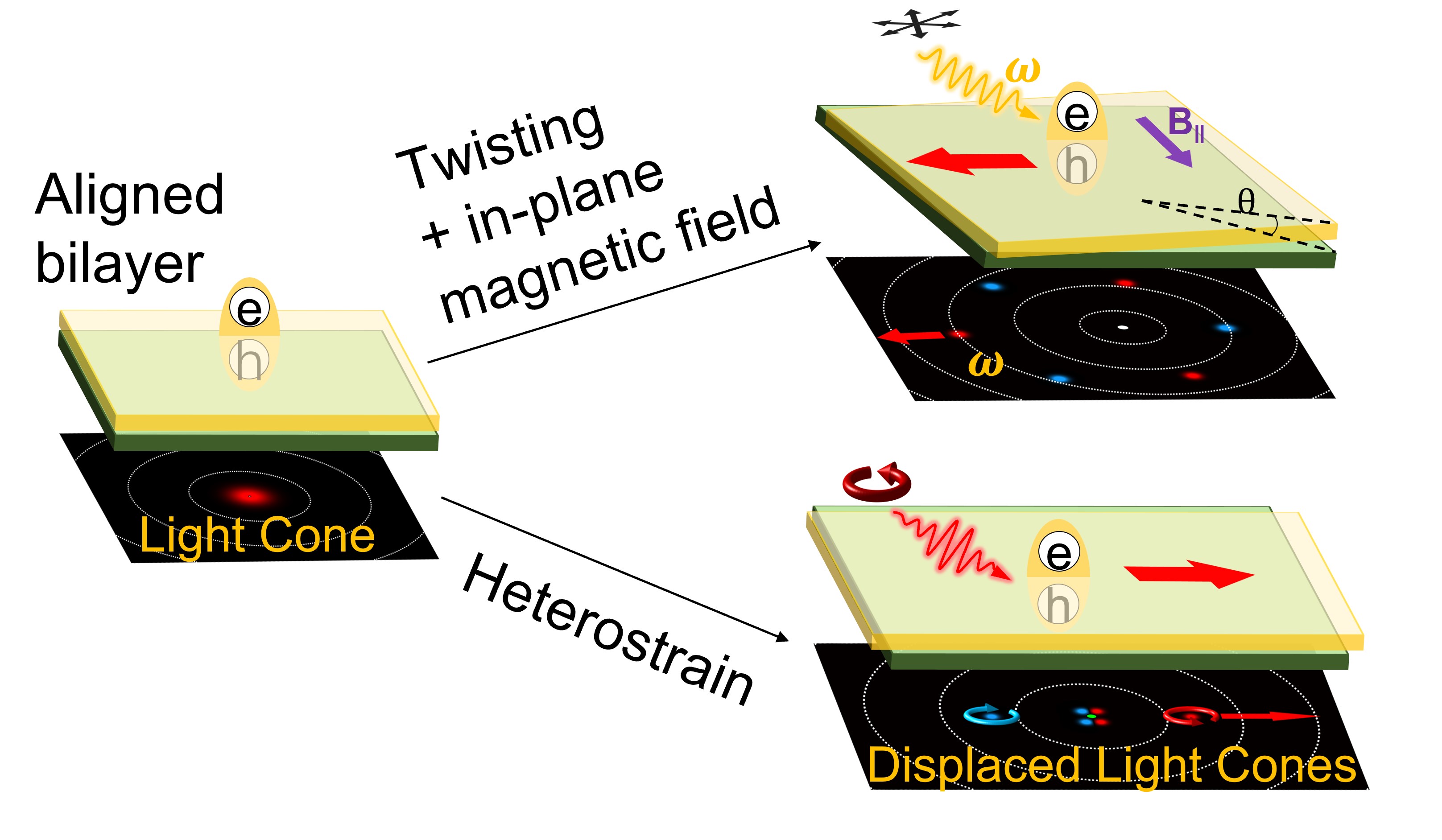}

\end{tocentry}

\begin{abstract}
  An anomalous magneto-optical spectrum is discovered for dipolar valley excitons in twisted double layer transition metal dichalcogenides, where in-plane magnetic field induces a sizable multiplet splitting of exciton states inside the light cone. Chiral dispersions of the split branches
make possible efficient optical injection of unidirectional exciton current.
We also find an analog effect with a modest heterostrain replacing the magnetic field for introducing large splitting and chiral dispersions in the light cone. Angular orientation of photo-injected exciton flow can be controlled by strain, with left-right unidirectionality selected by circular polarisation.
\end{abstract}


Transition metal dichalcogenides (TMDs) have become an exciting platform to explore optoelectronics and magneto-optics, with the appealing optical properties of tightly bound excitons formed in the degenerate valleys at Brillouin zone (BZ) corners. Exciton in a monolayer has strong light coupling with valley contrasted circularly polarized selection rules~\cite{wang2012electronics,xu2014spin,yu2015valley,mak2016photonics,wang2018colloquium}. 
In multilayer, interlayer exciton (IX) can also form with electron and hole separated to adjacent layers, acquiring long population and valley lifetimes at the cost of significant reduction in optical dipole. IX's permanent electrical dipole enables electrical tunability of its resonance, and strong exciton interactions for novel quantum many-body phenomena~\cite{fogler2014high}. 

In the general presence of twisting and lattice mismatch, IX's canonical and kinematic momenta become different, as BZ corners of the electron layer are displaced from those of the hole layer~\cite{yu2018brightened}. 
The `light cone' region, defined by the conservation of canonical momentum in the light coupling, furcates to six branches centered at finite kinematic momenta~\cite{yu2015anomalous}. 
This underlies the peculiar optical properties of IX discovered at small twist angles, where pronounced periodic potential in the long-wavelength moir\'e leads to exciton trapping and formation of mini-bands~\cite{yu2017moire,wu2017topological,tran2019evidence,seyler2019signatures,jin2019observation,zheng2021twist}.
The moir\'e potential can be smoothened by a thin BN spacer~\cite{shimazaki2020strongly,sun2022excitonic} or at large twist angle, while in such limit the small optical dipole and finite kinetic energy in the light cone tend to make IX optically inactive. 

When intra- and inter-layer excitons are brought close in resonance, they can hybridize via interlayer hopping of electron/hole, as observed in various heterobilayers~\cite{alexeev2019resonantly,hsu2019tailoring,zhang2020twist,tang2021tuning},
and markedly in MoSe$_2$/BN/MoSe$_2$ with monolayer BN spacer~\cite{shimazaki2020strongly}.
Such hybrid excitons that inherit both the large optical dipole and permanent electrical dipole are favourable for excitonic quantum phenomena with optical and electrical addressability~\cite{yu2021luminescence}.

Magnetic control of valley excitons is also under active exploration. Because of the spin-valley locking, carriers in TMDs have strongly anisotropic g-factors in the out-of-plane direction only. Magneto-optical study of valley excitons has been focused on the sizable Zeeman splitting~\cite{macneill2015breaking,li2014valley,aivazian2015magnetic,srivastava2015valley,seyler2019signatures,baek2020highly}
and Landau level formation~\cite{wang2017valley,liu2020landau,li2020phonon,wang2020observation,li2020spontaneous,smolenski2019interaction,klein2021controlling}
in out-of-plane magnetic field $\mathbf{B}_\perp$.
Exciton with a finite kinematic momentum $\mathbf{Q}$ can also respond to magnetic field $\mathbf{B}$ through the magneto-Stark effect, where $\mathbf{Q} \times \mathbf{B}$ acts as effective electric field in the moving frame that couples to the electrical dipole~\cite{thomas1961amagneto,hopfield1961fine,gorkov1968contribution,chang2002quantum}.
When the light cone is centered at zero $\mathbf{Q}$, observable manifestation of this effect is limited, either by utilizing the tiny $\mathbf{Q}$ at the light cone edge~\cite{karin2016giant}, or in nonlinear optical processes~\cite{farenbruch2020magneto,lafrentz2013magneto}.

\begin{figure}
	\centering
	\includegraphics[width=0.48\textwidth]{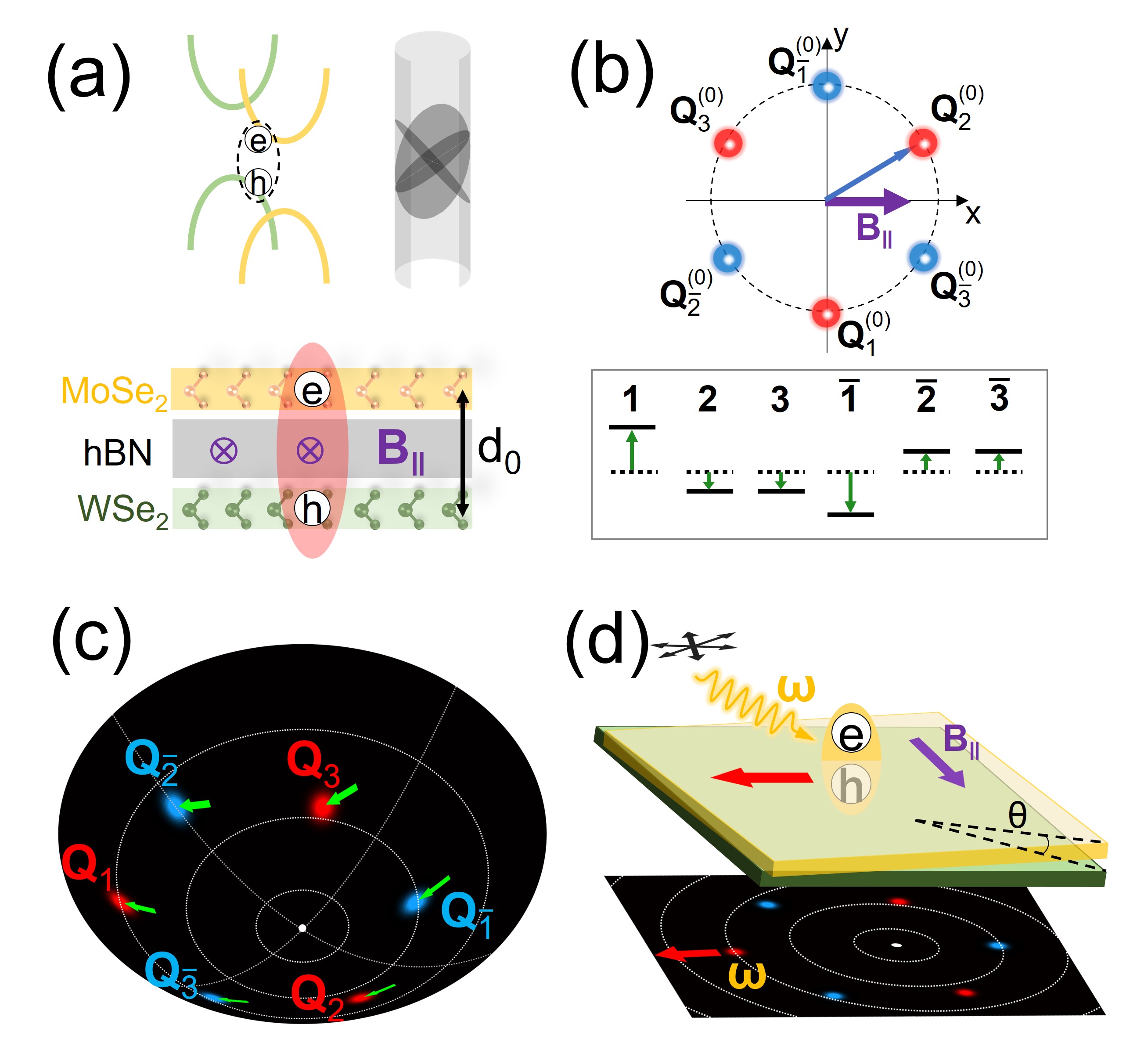}
	\caption{(a) Schematic of IX in double-layer TMDs with $\mathbf{B}_\parallel$. On the top shows IX dispersion in the light cone, which has multiple branches with three-fold rotational symmetry. (b) Upper: by twisting, the light cone region multi-furcates into six branches centered at finite kinematic momenta $\mathbf{Q}^{(0)}_{i/\bar{i}}$. Lower: distinct magneto-Stark shift of each branch, $\propto \mathbf{Q}^{(0)} \times \mathbf{B}$, leads to multiplet splitting in light cone. (c) The magneto-Stark effect can be described as a change of the kinematic momenta of the light cone branches by $\mathbf{B}_\parallel$, while the same momentum change (green arrows) corresponds to distinct energy shift at different places of the kinetic energy dispersion. The dotted rings denote equal-energy contours. (d) The multiplet splitting allows injection of uni-directional exciton current (red arrow), by resonant excitation at selected branch. }
	\label{fig_introduction}    
\end{figure}

For the dipolar excitons in twisted TMDs, we discover here the anomalous magneto-optical response in the regime where moir\'e potential is smoothened at large twist angles or by a BN spacer. With the light cone branches furcated to six finite kinematic momenta (Fig.~\ref{fig_introduction}b), the magneto-Stark effect now manifests as a sizable multiplet splitting proportional to the in-plane magnetic field $\mathbf{B}_\parallel$. The g-tensor is determined by the interference of the magneto-Stark shift from $\mathbf{B}_\parallel$ and the Zeeman shift due to $\mathbf{B}_\perp$, and the axes of anisotropy point in distinct angles for the six light cone branches. When dipolar excitons become bright through hybridization with intralayer exciton, resonant excitation of the split branches of chiral dispersions makes possible efficient injection of unidirectional exciton current.
The magneto-Stark effect can be alternatively interpreted from the perspective of layer-contrasted vector potential. Along this line, we show that heterostrain, effectively described as layer-contrasted pseudo vector potential, can also result in sizable multiplet splitting and chiral dispersions in the light cone. This makes possible a chiral light-matter interface without the need of magnetic field, where the left-right directionality of injected exciton current is controlled by circular polarisation. 

\textit{Magneto-Stark effect of IX at the misaligned valleys.} -- 
In the considered regime with negligible moir\'e potential, the Hamiltonian for an interlayer electron-hole pair in the envelope function approximation is,
\begin{eqnarray}
\hat{H} &=& \frac{\hbar^2}{2m_e} [-i\frac{\partial}{\partial \mathbf{r}_e} - \mathbf{K}_t + \mathbf{A}^t(\mathbf{r}_e)]^2 + \frac{\hbar^2}{2m_h}[-i\frac{\partial}{\partial \mathbf{r}_h} + \mathbf{K}_b - \mathbf{A}^b(\mathbf{r}_h)]^2 \\
&+& V(|\mathbf{r}_e-\mathbf{r}_h|) \notag,
\end{eqnarray}
where $\mathbf{r}_e(\mathbf{r}_h)$ is the in-plane coordinate of electron (hole), and $V$ is the interlayer Coulomb interaction, $\mathbf{A}^{t/b}$ is the vector potential of in-plane magnetic field $\mathbf{B}_\parallel$, and $\mathbf{K}_{t/b}$ the misaligned Dirac point at the two layers. $\frac{e}{\hbar}=1$ is used for brevity. With the ultrastrong Coulomb binding in TMDs, the electron-hole relative motion is unaffected in the range of $\mathbf{B}_\parallel$ considered, and remains separable from the center-of-mass (COM) motion (with coordinate $\mathbf{R} \equiv \frac{m_e\mathbf{r}_e+m_h\mathbf{r}_h}{M}$, $M \equiv m_e+m_h$). The {\it canonical} momentum operator for exciton's COM motion is then: $\hat{\mathbf{\Pi}} = \hat{\mathbf{Q}} - \mathbf{Q}^{(0)} - d_0 \mathbf{B}_\parallel \times \hat{\mathbf{z}}$~\cite{herold1981the,gorkov1968contribution}, where $\hat{\mathbf{Q}} \equiv \frac{i}{\hbar}[\hat{H},\mathbf{R}] $ is the operator for {\it kinematic} momentum,
$d_0$ the interlayer distance (Fig.~\ref{fig_introduction}a), and $\mathbf{Q}^{(0)} = \mathbf{K}_b-\mathbf{K}_t$. One can show that $[\hat{\mathbf{\Pi}},\hat{H}] = 0$, and $[\hat{\Pi}_{x},\hat{\Pi}_{y}] = 0$. It is this canonical momentum that need to be matched in the interconversion with photon, which defines the light cone region centered at its zero eigenvalue. 

In the absence of $\mathbf{B}_\parallel$, the kinematic and canonical momenta simply differ by $\mathbf{Q}^{(0)}$. Twisting and lattice mismatch therefore lead to finite kinetic energy of valley excitons in the light cone region. The freedom in expressing the Dirac points by adding reciprocal vectors of the two layers leads to multiple values of $\mathbf{Q}^{(0)}$ separated by the moir\'e reciprocal vectors, corresponding to multiple branches (or mini-bands) inside the light cone (see Fig.~\ref{fig_introduction}a top right)~\cite{yu2015anomalous}. Concerning the optical response and hybridization with intralayer exciton, 
the active branches are only those with the smallest $\mathbf{Q}^{(0)}$, three for K valley exciton ($\mathbf{Q}^{(0)}_{1,2,3}$) and their time-reversal ($\mathbf{Q}^{(0)}_{\bar{1},\bar{2},\bar{3}}$) for -K valley (see Fig.~\ref{fig_introduction}b). 

$\mathbf{B}_\parallel$ changes the relation between the canonical and kinematic momenta, consequently the centers of the light cone branches are shifted to kinematic momenta 
\begin{equation}
\mathbf{Q}_{i/\bar{i}}= \mathbf{Q}^{(0)}_{i/\bar{i}} + d_0 \mathbf{B}_\parallel \times \hat{\mathbf{z}}
\end{equation}
and the corresponding kinetic energies $E_{i/\bar{i}} = \frac{\hbar^2 \mathbf{Q}_{i/\bar{i}}^2}{2M}$ become  
\begin{eqnarray}
E_{i/\bar{i}} \approx \frac{\hbar^2 \mathbf{Q}_{i/\bar{i}}^{(0)^2}}{2M} + \frac{\hbar^2}{M} ( \mathbf{Q}_{i/\bar{i}}^{(0)} \times \mathbf{B}_\parallel ) \cdot d_0\hat{\mathbf{z}}.
\label{eq_kineticenergy}
\end{eqnarray}
where we keep only the linear response. The last term is the magneto-Stark effect which has a new manifestation at the misaligned valleys. First, the magnitude of the magneto-Stark shift gets significantly amplified by the large value of $\mathbf{Q}^{(0)}$ proportional to the twisting and lattice mismatch. Second, the different $\mathbf{Q}^{(0)}$ vectors of the multiple branches in the light cone allow distinct magnetic control of their energy (Fig.~\ref{fig_introduction}b-c). 
The resultant multiplet splitting, controlled by the in-plane angle of $\mathbf{B}_\parallel$, can be exploited for injection of unidirectional exciton current by resonant excitation (Fig.~\ref{fig_introduction}d). 

\begin{figure}
	\centering
	\includegraphics[width=0.48\textwidth]{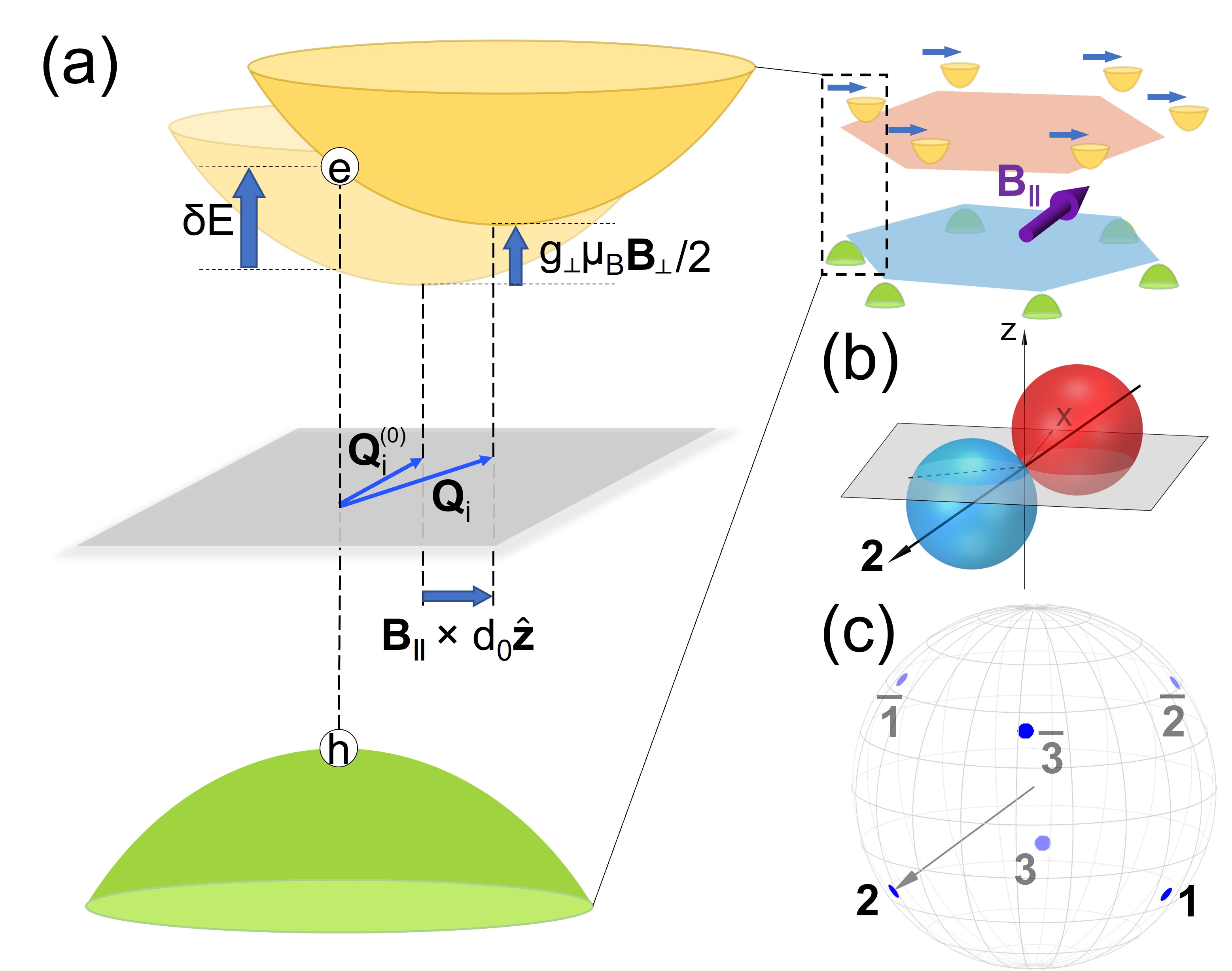}
	
	\caption{(a) Schematic of magneto-Stark shift of IX by $\mathbf{B}_{\parallel}$ and Zeeman shift by $\mathbf{B}_{\perp}$. $\mathbf{B}_{\parallel}$ adds an extra displacement of the electron valleys from the hole valleys (blue horizontal arrows). For an exciton in the light cone, its energy shift $\delta E$ by the field is the sum of the Zeeman and magneto-Stark shifts. (b) $\delta E$ of light cone branch $i=2$ as function of magnetic field direction. The radial length from coordinate origin to the surface of spheres gives the magnitude of $\delta E$ and blue/red color indicates positive/negative sign. The solid arrow denotes the axis of anisotropy, pointing in the field direction of maximum blue shift. (c) Axes of anisotropy for the six branches in light cone are indicated by the blue spots on a unit sphere.}
	\label{fig_3dzeeman}
\end{figure}

\textit{Anomalous g-tensor and magnetic anisotropy.} -- Because of the spin-valley locking in TMDs that separate the time reversal pair of spin up and down states to the two valleys, 
the magnetic splitting of TMDs excitons can only be in the out-of-plane direction. The Zeeman splitting by out-of-plane magnetic field $\mathbf{B}_\perp$ is known to have a sizable g-factor~\cite{aivazian2015magnetic}. For example, IX in the R-stacking $\text{MoSe}_2/\text{WSe}_2$ has a measured $g_\perp = 6.72$~\cite{seyler2019signatures}.
We find that the magneto-Stark effect at the misaligned valleys can have a comparable g-factor, which, together with the Zeeman effect, determines an unusual magnetic anisotropy.

In a general magnetic field with both in-plane and out-of-plane components, the magneto-shift of the IX eigenenergies in the light cone are
\begin{eqnarray}
\delta E_{i/\bar{i}} &=& \frac{\hbar^2}{M} ( \mathbf{Q}_{i/\bar{i}}^{(0)} \times \mathbf{B}_\parallel ) \cdot d_0\hat{\mathbf{z}} \pm \frac{1}{2}g_{\perp}\mu_B B_{\perp}.
\end{eqnarray}
Expressed as $\delta E = \mu_B \mathbf{B} \cdot \stackrel{\leftrightarrow}{g} \cdot \mathbf{S}$, the g-tensor for K valley states has the form
\begin{eqnarray}
    \stackrel{\leftrightarrow}{g}_i = g^0_i
    \begin{pmatrix}
    0 & 0 & -\text{sin} \, \phi_i\\
    0 & 0 & \text{cos} \, \phi_i\\
    0 & 0 & \text{tan} \, \xi_i\\
    \end{pmatrix}.
    \label{eq_gtensor}
\end{eqnarray}
$g^0_i = \frac{2m_0}{M}Q_i^{(0)} d_0$ is an effective g-factor with $m_0$ the free electron mass, and $\phi_i$ is the azimuthal angle of $\mathbf{Q}_i^{(0)}$. $\xi_i \equiv \text{tan}^{-1}(\frac{g_\perp}{2g_i^0})$ describes the relative strength of the in-plane and out-of-plane field effects. For the -K valley states, $\stackrel{\leftrightarrow}{g}_{\bar{i}} = -\stackrel{\leftrightarrow}{g}_i$. 

At fixed field magnitude, the polar plot of $\delta E_i$ as a function of field direction exhibits two tangent spheres, as shown in Fig.~\ref{fig_3dzeeman}b for the example of $5^\circ$ twisted MoSe$_2$/WSe$_2$, with exciton mass $M=0.84 m_0$, where a monolayer BN spacer leads to an electric dipole of $d_0=2$ nm. 
The maximum shift is reached along a field direction significantly tilted from the $z$-axis denoted by the arrow (Fig.~\ref{fig_3dzeeman}b). 
The six IX branches in the light cone are distinguished by their different axes of magnetic anisotropy (Fig.~\ref{fig_3dzeeman}c), which leads to a rich multiplet splitting pattern controlled by the field direction. Between a time-reversal pair of branches $i$ and $\bar{i}$, a maximum splitting is $0.78$ meV/T for the example shown. 
The polar angle of this axis of anisotropy represents the relative strength between the magneto-Stark effect and Zeeman effect. If the axis approaches the equator (pole), the magneto-Stark (Zeeman) effect dominates. 
With the increase of twist angle, the magneto-Stark effect is enhanced and the axes of anisotropy get tilted towards the equator.

\textit{Light cone splitting by the pseudo-vector potential from strain.} -- The magneto-Stark effect can also be understood from the perspective of layer-contrasted vector potential that adds to the displacement between the electron and hole valleys (Fig.~\ref{fig_3dzeeman}a). Choosing the gauge $\mathbf{A}^t = \mathbf{B}_\parallel \times d_0 \mathbf{z}$ and $\mathbf{A}^b=0$, the electron valleys are shifted to $\mathbf{K}_t-\mathbf{B}_\parallel \times d_0 \hat{\mathbf{z}}$, while the hole valleys remain at $\mathbf{K}_b$. Thus, the kinematic momentum of the light cone states becomes $\mathbf{K}_b - (\mathbf{K}_t-\mathbf{B}_\parallel \times d_0 \hat{\mathbf{z}}) = \mathbf{Q}^{(0)}+\mathbf{B}_\parallel \times d_0 \hat{\mathbf{z}}$.
We note that strain effect also leads to pseudo-vector potentials that shift the valleys in momentum space, so a layer-dependent strain (heterostrain) may play a similar role as the in-plane magnetic field for lifting the degeneracy of the light cone branches. 

\begin{figure}
	\centering
	\includegraphics[width=0.48\textwidth]{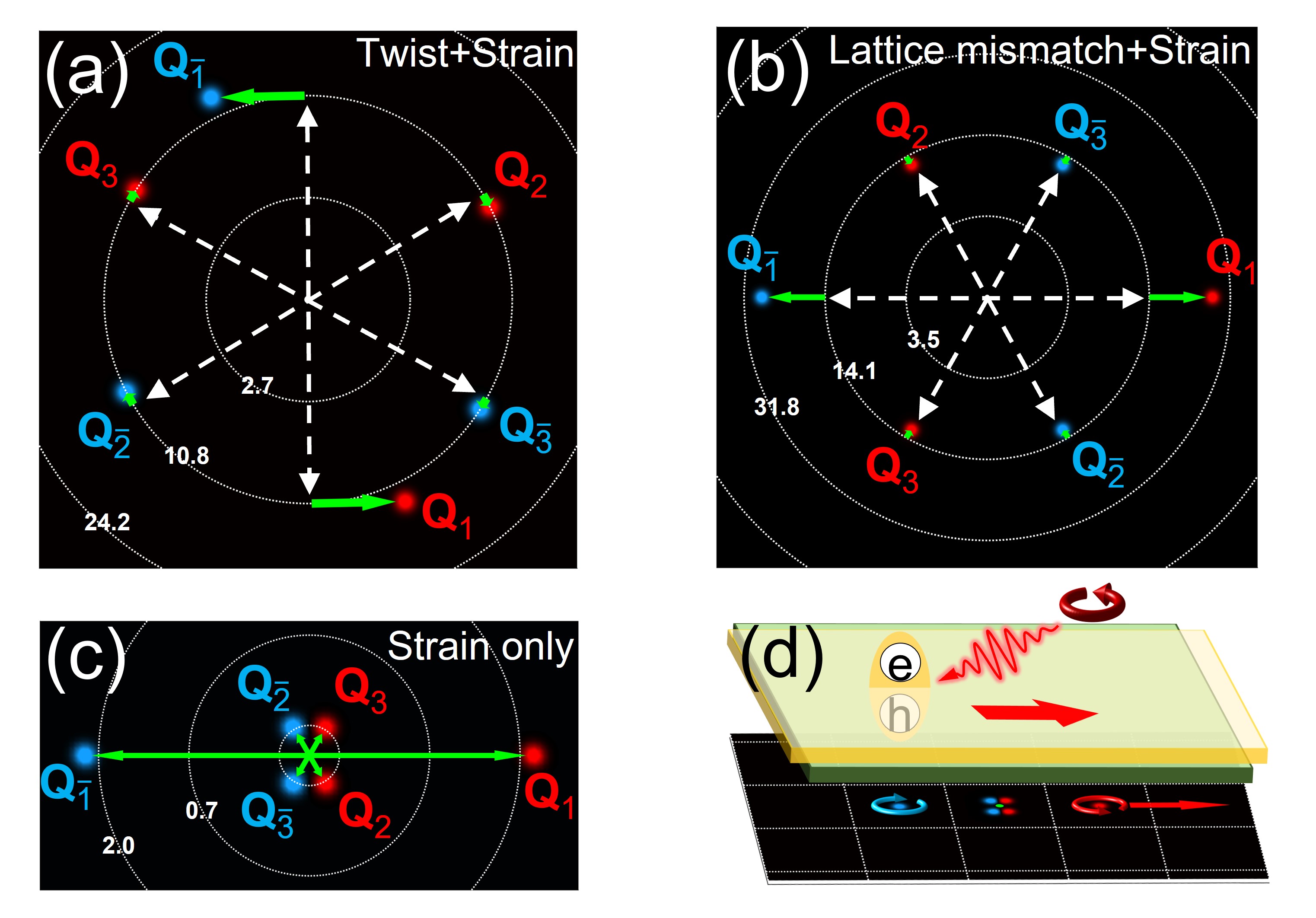}
	
	\caption{(a) The displacements (green arrows) in kinematic momenta of the six light cone branches by uniaxial heterostrain are perpendicular to $\mathbf{Q}_{i,\bar{i}}^{(0)}$ due to twisting (dashed arrow). The plot is for a $2^\circ$ twisting with $1\%$ heterostrain of Poisson's ratio 0.23. (b) The case of $4\%$ lattice mismatch, by the same heterostrain as in (a). (c) The special case of only heterostrain as in (a). Energy (in meV) is marked on the equal-energy contours. (d) Injection of unidirectional valley exciton current with large group velocity, by circularly polarized excitation.}
	\label{fig_strain}    
\end{figure}

Assuming strain is applied on the top (electron) layer. The electron valleys are now centered at $\mathbf{D}_e = (1 + \stackrel{\leftrightarrow}{\epsilon})^{-1} \mathbf{K}_t-\mathbf{A}_\text{S}$, where $ \stackrel{\leftrightarrow}{\epsilon}=\left(\begin{array}{cc} \epsilon_{xx} & \epsilon_{xy}  \\ \epsilon_{yx} & \epsilon_{yy} \\ \end{array}\right)$ is the strain tensor. $\mathbf{A}_\text{S} = \pm \beta K (\epsilon_{xx}-\epsilon_{yy},-2\epsilon_{xy})$ is a pseudo vector potential due to the strain modulation of the intralayer hopping, where time-reversal symmetry requires opposite sign at K and -K valleys. $\beta$ is a material-dependent parameter~\cite{amorim2016novel}, and $K\equiv |\mathbf{K}_t|$.
A modest heterostrain therefore displaces the kinematic momenta of the light cone branches by $d\mathbf{Q} \approx \stackrel{\leftrightarrow}{\epsilon}\mathbf{K}_t + \mathbf{A}_\text{S}$. Consider a uniaxial strain $ \stackrel{\leftrightarrow}{\epsilon}=\left(\begin{array}{cc} \epsilon & 0  \\ 0 & - \nu \epsilon \\ \end{array}\right)$.
Adopting Poisson's ratio $\nu = 0.23$ and $\beta = 0.49$ for MoSe$_2$~\cite{kang2013band,fang2018electronic}, Fig.~\ref{fig_strain}a-c show the strain induced displacements $d\mathbf{Q}$, for three exemplary cases respectively with $\theta = 2^{\circ}$ twisting (3a), $\eta = 4\%$ lattice mismatch (3b), and zero twisting and lattice mismatch (3c). 
$d\mathbf{Q}$ is approximately perpendicular to $\mathbf{Q}^{(0)}_i$ introduced by twisting, and parallel to $\mathbf{Q}^{(0)}_i$ from lattice mismatch. The strain induced kinetic energy change in the light cone is $\propto d\text{Q}^2$ in the former, and $\propto \text{Q}^{(0)}d\text{Q}$ in the latter case. Importantly, the change is different for the three light cone branches from each valley (see Fig.~\ref{fig_strain}), which are then split into a singlet ($i=1$) and a  doublet ($i=2,3$). The magnitude of the strain induced splitting between the two groups is
\begin{eqnarray}
\delta E_1 - \delta E_{2,3} &=& \frac{\hbar^2 K^2}{2M}\biggl[\frac{3}{4} (1-\nu^2) \epsilon^2 + 3 \beta (1+\nu) \epsilon^2 \notag\\
&+&\frac{3}{2} (1+2\beta) (1+\nu) \eta \epsilon + \frac{\sqrt{3}}{2}(1-2\beta)(1+\nu)\theta\epsilon \biggr].
\label{eq_splitting}
\end{eqnarray}
In the case of Fig.~\ref{fig_strain}a and Fig.~\ref{fig_strain}c, the splitting is about 2 meV at $\epsilon = 1\%$, contributed by the two $\epsilon^2$ terms. Lattice mismatch can dramatically amplify the strain induced splitting through the third term $\propto \eta \epsilon$. With $\eta = 4\%$ (e.g. for WS$_2$/WSe$_2$), $\epsilon= 1\%$ strain can introduce a sizable splitting of 15 meV (Fig.~\ref{fig_strain}b).

\textit{Chiral light-matter interface through exciton hybridization.} -- The degeneracy-lifted light cone branches due to magnetic field or heterostrain can be utilized to realize chiral light-matter interface. These exciton states carry finite kinematic momenta, with distinct moving directions at the six branches.
$\mathbf{B}_\parallel$ breaks the time-reversal and three-fold rotational symmetry, and an individual light cone branch $i$ can be frequency selected in resonant excitation, for injection of unidirectional flow of dipolar excitons with group velocity $\mathbf{Q}_i$ (Fig.~\ref{fig_introduction}d). Heterostrain preserves the time reversal symmetry, so the split branches have opposite chirality at the two valleys. This can be exploited to inject either pure valley current, or unidirectional exciton dipole current through the valley selective excitation (Fig.~\ref{fig_strain}d). While IX has small optical dipole by itself~\cite{yu2015anomalous}, through hybridization with intralayer exciton~\cite{shimazaki2020strongly,alexeev2019resonantly,hsu2019tailoring,zhang2020twist,tang2021tuning}, an efficient chiral light-matter interface can be realized.

\begin{figure}
	\centering
	\includegraphics[width=0.48\textwidth]{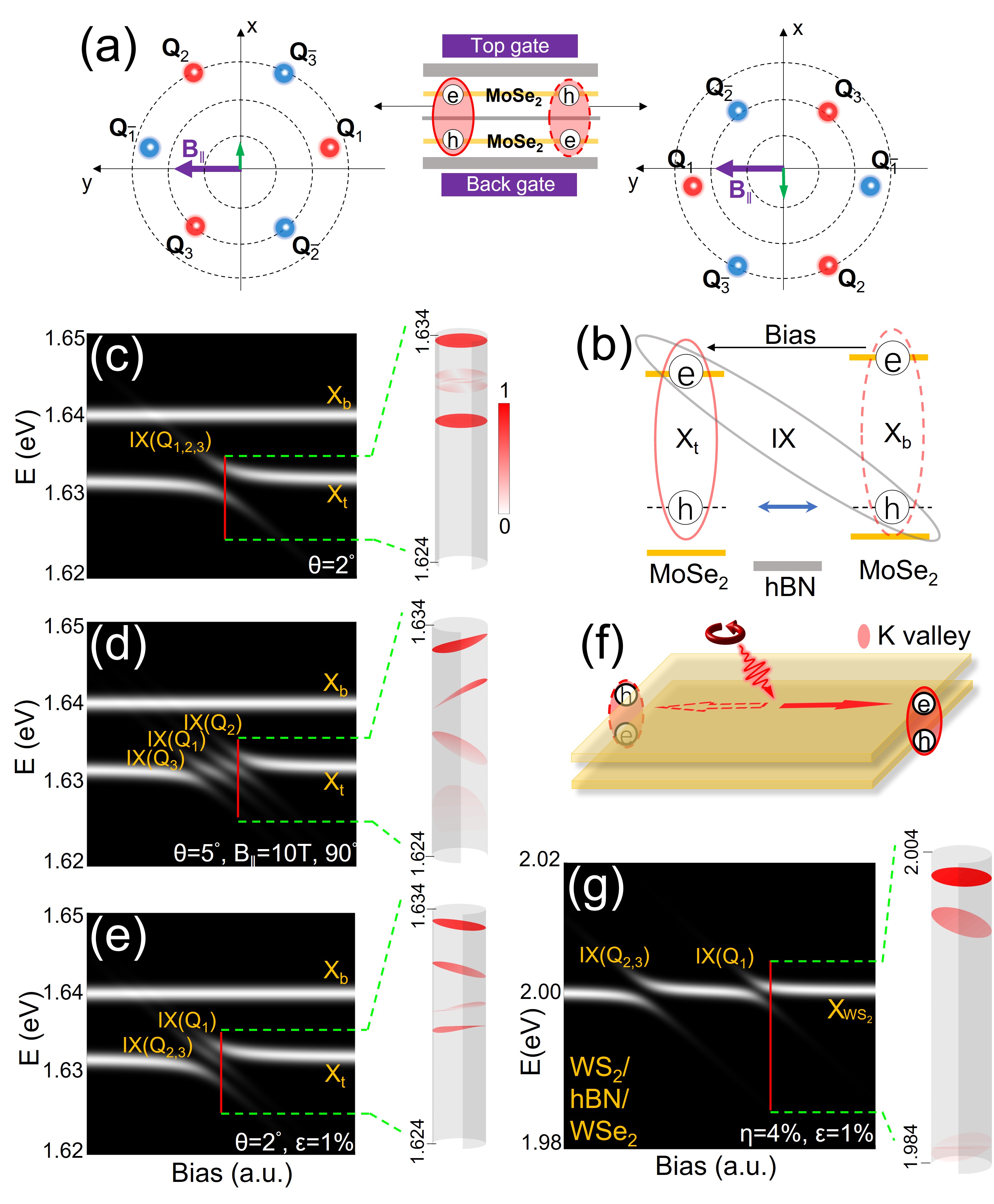}
	
	\caption{(a) MoSe$_2$ double-layer with a BN spacer, hosting two species of IX with opposite dipoles and therefore opposite magneto-Stark shifts. (b) At certain interlayer bias, the IX shown can hybridize with intralayer exciton in top layer ($X_t$) through the hole hopping. (c) Simulated reflectance spectrum (left); and exciton dispersions in the light cone (right) at the interlayer bias marked by the red line, at $2^\circ$ twisting, without magnetic field or strain. The brightness, i.e. optical dipole strength, is colour-coded on the dispersions. (d) Similar plot for $5^\circ$ twisting, in $\mathbf{B}_\parallel = 10$ T, and (e) for $2^\circ$ twist angle, with $1\%$ heterostrain (same as Fig.~\ref{fig_strain}a). Chiral branches with pronounced optical dipole are obtained, which realize an efficient chiral interface for photo-injection of unidirectional exciton current. In (c-e), the interlayer hopping strength 1 meV is taken. (f) Schematic of injection by circularly polarized light. The solid and dashed arrow indicate the flow direction for the two IX species respectively. (g) Hybrid exciton in WS$_2$/WSe$_2$ double-layer with $1\%$ heterostrain. Both the splitting and group velocity of chiral branches are amplified by the large lattice mismatch.}
	\label{fig_hybridization}    
\end{figure}

Using the exemplary system of R-stacked MoSe$_2$/hBN/MoSe$_2$ where gate tunable hybridization is observed~\cite{shimazaki2020strongly}, we show the hybrid excitons inherit both the chiral dispersion from the anomalous magneto or strain splitting from its IX component, and the sizable optical dipole with valley selectivity from its intralayer component. The analysis on H-stacked bilayers is available in the Supporting Information.  
Experiment has found that hybridization is through the hole hopping, which has pronounced amplitude even in the presence of a monolayer BN spacer~\cite{shimazaki2020strongly}, attributed to the band alignment of BN and MoSe$_2$ (Fig.~\ref{fig_hybridization}b). In our calculations, we have assumed negligible electron interlayer tunneling across the hBN spacer, and the exciton layer hybridization is through the hole tunneling~\footnote{If electron tunneling is present in the exciton layer hybridization, the anti-crossing behaviors in the spectrum can be quantitatively different.}.
Fig.~\ref{fig_hybridization}c shows the calculated reflection spectrum in the absence of magnetic field and heterostrain. The hole interlayer hopping is taken to be 1 meV, which produces a similar anti-crossing pattern as observed~\cite{shimazaki2020strongly}. The energy difference between intralayer branches $X_t$ and $X_b$ is also adopted from the experiment.
The brightness corresponds to the oscillator strength of the exciton state at the center of light cone (normal incidence). 
In a large neighborhood of the anti-crossing, the dipolar exciton acquires sizable oscillator strength for efficient optical excitation, while retaining the electrical dipole as reflected by the electrical Stark shift. At a selected bias value (red line), the dispersions of the exciton states are shown in the light cone, which retains the three-fold rotational symmetry. 

Fig.~\ref{fig_hybridization}d shows the case with a $\mathbf{B}_\parallel = 10$ T in-plane magnetic field. The splitting of the light cone branches by the magneto-Stark effect is clearly visible in the reflective spectrum. Correspondingly, the dispersions becomes chiral in the light cone. At the selected bias value, the top branch shown in Fig.~\ref{fig_hybridization}d is dominantly the intralayer $X_t$, which also acquires a finite group velocity through a small IX component. The second branch is dominantly an IX with large group velocity, while featuring a sizable optical dipole that is a significant fraction of that of $X_t$. Unidirectional current injection can be efficiently realized for both branches. The same functionality can also be realized by heterostrain as shown in Fig.~\ref{fig_hybridization}e. In Fig.~\ref{fig_hybridization}d-e, for clarity, only the K valley states are shown in the light cone. Experimentally, one can also selectively examine one valley using circularly polarized light, where the spectra can be simpler for analyzing the field angle dependence (c.f. the Supporting Information). As the optical dipole is from the intralayer exciton, it has $\sigma+$ ($\sigma-$) polarisation for the K (-K) valley states, which allows valley selection. 

The strain effect is expected to be much more pronounced in WS$_2$/WSe$_2$ double layer which also hosts IX that is close in energy with the monolayer exciton in WS$_2$~\cite{tang2021tuning}.
From Eq.~(\ref{eq_splitting}), it is clear that both the splitting in the light cone and the group velocity of the chiral branch are significantly amplified by the large lattice mismatch (Fig.~\ref{fig_hybridization}g).
The MoSe$_2$ double layer, on the other hand, hosts two species of IX with opposite electrical dipole (Fig.~\ref{fig_hybridization}a), which can be gate tuned into resonance with the intralayer excitons at distinct bias range. The two species of dipole excitons feature opposite magneto-Stark shift, and opposite chirality of the split branches, as schematically shown in Fig.~\ref{fig_hybridization}f. This may be exploited to realize a light-matter interface with a chirality that can be switched both by light polarisation and electrical bias. 

\begin{acknowledgement}
The work is supported by a grant (AoE/P-701/20) and a fellowship award (HKU SRFS2122-7S05) from the Research Grant Council of Hong Kong SAR, and by the National Key R\&D Program of China (2020YFA0309600). W. Y. acknowledges support by Tencent Foundation.
\end{acknowledgement}


\begin{suppinfo}
The Supporting Information is available free of charge at http://pubs.acs.org.

I. Field angle dependence and the two-valley analysis of the absorption spectrum of layer hybridized excitons. II. Analysis of the hybridized excitons on H-stacked bilayers.
\end{suppinfo}

\bibliography{reference}

\providecommand{\latin}[1]{#1}
\makeatletter
\providecommand{\doi}
  {\begingroup\let\do\@makeother\dospecials
  \catcode`\{=1 \catcode`\}=2 \doi@aux}
\providecommand{\doi@aux}[1]{\endgroup\texttt{#1}}
\makeatother
\providecommand*\mcitethebibliography{\thebibliography}
\csname @ifundefined\endcsname{endmcitethebibliography}
  {\let\endmcitethebibliography\endthebibliography}{}
\begin{mcitethebibliography}{45}
\providecommand*\natexlab[1]{#1}
\providecommand*\mciteSetBstSublistMode[1]{}
\providecommand*\mciteSetBstMaxWidthForm[2]{}
\providecommand*\mciteBstWouldAddEndPuncttrue
  {\def\EndOfBibitem{\unskip.}}
\providecommand*\mciteBstWouldAddEndPunctfalse
  {\let\EndOfBibitem\relax}
\providecommand*\mciteSetBstMidEndSepPunct[3]{}
\providecommand*\mciteSetBstSublistLabelBeginEnd[3]{}
\providecommand*\EndOfBibitem{}
\mciteSetBstSublistMode{f}
\mciteSetBstMaxWidthForm{subitem}{(\alph{mcitesubitemcount})}
\mciteSetBstSublistLabelBeginEnd
  {\mcitemaxwidthsubitemform\space}
  {\relax}
  {\relax}

\bibitem[Wang \latin{et~al.}(2012)Wang, Kalantar-Zadeh, Kis, Coleman, and
  Strano]{wang2012electronics}
Wang,~Q.~H.; Kalantar-Zadeh,~K.; Kis,~A.; Coleman,~J.~N.; Strano,~M.~S.
  Electronics and optoelectronics of two-dimensional transition metal
  dichalcogenides. \emph{Nat. Nanotechnol.} \textbf{2012}, \emph{7},
  699--712\relax
\mciteBstWouldAddEndPuncttrue
\mciteSetBstMidEndSepPunct{\mcitedefaultmidpunct}
{\mcitedefaultendpunct}{\mcitedefaultseppunct}\relax
\EndOfBibitem
\bibitem[Xu \latin{et~al.}(2014)Xu, Yao, Xiao, and Heinz]{xu2014spin}
Xu,~X.; Yao,~W.; Xiao,~D.; Heinz,~T.~F. Spin and pseudospins in layered
  transition metal dichalcogenides. \emph{Nat. Phys.} \textbf{2014}, \emph{10},
  343--350\relax
\mciteBstWouldAddEndPuncttrue
\mciteSetBstMidEndSepPunct{\mcitedefaultmidpunct}
{\mcitedefaultendpunct}{\mcitedefaultseppunct}\relax
\EndOfBibitem
\bibitem[Yu \latin{et~al.}(2015)Yu, Cui, Xu, and Yao]{yu2015valley}
Yu,~H.; Cui,~X.; Xu,~X.; Yao,~W. {Valley excitons in two-dimensional
  semiconductors}. \emph{Natl. Sci. Rev.} \textbf{2015}, \emph{2}, 57--70\relax
\mciteBstWouldAddEndPuncttrue
\mciteSetBstMidEndSepPunct{\mcitedefaultmidpunct}
{\mcitedefaultendpunct}{\mcitedefaultseppunct}\relax
\EndOfBibitem
\bibitem[Mak and Shan(2016)Mak, and Shan]{mak2016photonics}
Mak,~K.~F.; Shan,~J. Photonics and optoelectronics of 2D semiconductor
  transition metal dichalcogenides. \emph{Nat. Photonics} \textbf{2016},
  \emph{10}, 216--226\relax
\mciteBstWouldAddEndPuncttrue
\mciteSetBstMidEndSepPunct{\mcitedefaultmidpunct}
{\mcitedefaultendpunct}{\mcitedefaultseppunct}\relax
\EndOfBibitem
\bibitem[Wang \latin{et~al.}(2018)Wang, Chernikov, Glazov, Heinz, Marie, Amand,
  and Urbaszek]{wang2018colloquium}
Wang,~G.; Chernikov,~A.; Glazov,~M.~M.; Heinz,~T.~F.; Marie,~X.; Amand,~T.;
  Urbaszek,~B. Colloquium: Excitons in atomically thin transition metal
  dichalcogenides. \emph{Rev. Mod. Phys.} \textbf{2018}, \emph{90},
  021001\relax
\mciteBstWouldAddEndPuncttrue
\mciteSetBstMidEndSepPunct{\mcitedefaultmidpunct}
{\mcitedefaultendpunct}{\mcitedefaultseppunct}\relax
\EndOfBibitem
\bibitem[Fogler \latin{et~al.}(2014)Fogler, Butov, and
  Novoselov]{fogler2014high}
Fogler,~M.~M.; Butov,~L.~V.; Novoselov,~K.~S. High-temperature superfluidity
  with indirect excitons in van der Waals heterostructures. \emph{Nat. Commun.}
  \textbf{2014}, \emph{5}, 4555\relax
\mciteBstWouldAddEndPuncttrue
\mciteSetBstMidEndSepPunct{\mcitedefaultmidpunct}
{\mcitedefaultendpunct}{\mcitedefaultseppunct}\relax
\EndOfBibitem
\bibitem[Yu \latin{et~al.}(2018)Yu, Liu, and Yao]{yu2018brightened}
Yu,~H.; Liu,~G.-B.; Yao,~W. Brightened spin-triplet interlayer excitons and
  optical selection rules in van der Waals heterobilayers. \emph{2D Mater.}
  \textbf{2018}, \emph{5}, 035021\relax
\mciteBstWouldAddEndPuncttrue
\mciteSetBstMidEndSepPunct{\mcitedefaultmidpunct}
{\mcitedefaultendpunct}{\mcitedefaultseppunct}\relax
\EndOfBibitem
\bibitem[Yu \latin{et~al.}(2015)Yu, Wang, Tong, Xu, and Yao]{yu2015anomalous}
Yu,~H.; Wang,~Y.; Tong,~Q.; Xu,~X.; Yao,~W. Anomalous Light Cones and Valley
  Optical Selection Rules of Interlayer Excitons in Twisted Heterobilayers.
  \emph{Phys. Rev. Lett.} \textbf{2015}, \emph{115}, 187002\relax
\mciteBstWouldAddEndPuncttrue
\mciteSetBstMidEndSepPunct{\mcitedefaultmidpunct}
{\mcitedefaultendpunct}{\mcitedefaultseppunct}\relax
\EndOfBibitem
\bibitem[Yu \latin{et~al.}(2017)Yu, Liu, Tang, Xu, and Yao]{yu2017moire}
Yu,~H.; Liu,~G.-B.; Tang,~J.; Xu,~X.; Yao,~W. Moir{\'e} excitons: From
  programmable quantum emitter arrays to spin-orbit{\textendash}coupled
  artificial lattices. \emph{Sci. Adv.} \textbf{2017}, \emph{3}, e1701696\relax
\mciteBstWouldAddEndPuncttrue
\mciteSetBstMidEndSepPunct{\mcitedefaultmidpunct}
{\mcitedefaultendpunct}{\mcitedefaultseppunct}\relax
\EndOfBibitem
\bibitem[Wu \latin{et~al.}(2017)Wu, Lovorn, and MacDonald]{wu2017topological}
Wu,~F.; Lovorn,~T.; MacDonald,~A.~H. Topological Exciton Bands in Moir\'e
  Heterojunctions. \emph{Phys. Rev. Lett.} \textbf{2017}, \emph{118},
  147401\relax
\mciteBstWouldAddEndPuncttrue
\mciteSetBstMidEndSepPunct{\mcitedefaultmidpunct}
{\mcitedefaultendpunct}{\mcitedefaultseppunct}\relax
\EndOfBibitem
\bibitem[Tran \latin{et~al.}(2019)Tran, Moody, Wu, Lu, Choi, Kim, Rai, Sanchez,
  Quan, Singh, Embley, Zepeda, Campbell, Autry, Taniguchi, Watanabe, Lu,
  Banerjee, Silverman, Kim, Tutuc, Yang, MacDonald, and Li]{tran2019evidence}
Tran,~K. \latin{et~al.}  Evidence for moir{\'e} excitons in van der Waals
  heterostructures. \emph{Nature} \textbf{2019}, \emph{567}, 71--75\relax
\mciteBstWouldAddEndPuncttrue
\mciteSetBstMidEndSepPunct{\mcitedefaultmidpunct}
{\mcitedefaultendpunct}{\mcitedefaultseppunct}\relax
\EndOfBibitem
\bibitem[Seyler \latin{et~al.}(2019)Seyler, Rivera, Yu, Wilson, Ray, Mandrus,
  Yan, Yao, and Xu]{seyler2019signatures}
Seyler,~K.~L.; Rivera,~P.; Yu,~H.; Wilson,~N.~P.; Ray,~E.~L.; Mandrus,~D.~G.;
  Yan,~J.; Yao,~W.; Xu,~X. Signatures of moir{\'e}-trapped valley excitons in
  ${\mathrm{MoSe}}_{2}$/${\mathrm{WSe}}_{2}$ heterobilayers. \emph{Nature}
  \textbf{2019}, \emph{567}, 66--70\relax
\mciteBstWouldAddEndPuncttrue
\mciteSetBstMidEndSepPunct{\mcitedefaultmidpunct}
{\mcitedefaultendpunct}{\mcitedefaultseppunct}\relax
\EndOfBibitem
\bibitem[Jin \latin{et~al.}(2019)Jin, Regan, Yan, Utama, Wang, Zhao, Qin, Yang,
  Zheng, Shi, \latin{et~al.} others]{jin2019observation}
Jin,~C.; Regan,~E.~C.; Yan,~A.; Utama,~M. I.~B.; Wang,~D.; Zhao,~S.; Qin,~Y.;
  Yang,~S.; Zheng,~Z.; Shi,~S., \latin{et~al.}  Observation of moir{\'e}
  excitons in ${\mathrm{WSe}}_{2}$/${\mathrm{WS}}_{2}$ heterostructure
  superlattices. \emph{Nature} \textbf{2019}, \emph{567}, 76--80\relax
\mciteBstWouldAddEndPuncttrue
\mciteSetBstMidEndSepPunct{\mcitedefaultmidpunct}
{\mcitedefaultendpunct}{\mcitedefaultseppunct}\relax
\EndOfBibitem
\bibitem[Zheng \latin{et~al.}(2021)Zheng, Zhai, and Yao]{zheng2021twist}
Zheng,~H.; Zhai,~D.; Yao,~W. Twist versus heterostrain control of optical
  properties of moir{\'{e}} exciton minibands. \emph{2D Mater.} \textbf{2021},
  \emph{8}, 044016\relax
\mciteBstWouldAddEndPuncttrue
\mciteSetBstMidEndSepPunct{\mcitedefaultmidpunct}
{\mcitedefaultendpunct}{\mcitedefaultseppunct}\relax
\EndOfBibitem
\bibitem[Shimazaki \latin{et~al.}(2020)Shimazaki, Schwartz, Watanabe,
  Taniguchi, Kroner, and Imamo{\u{g}}lu]{shimazaki2020strongly}
Shimazaki,~Y.; Schwartz,~I.; Watanabe,~K.; Taniguchi,~T.; Kroner,~M.;
  Imamo{\u{g}}lu,~A. Strongly correlated electrons and hybrid excitons in a
  moir{\'e} heterostructure. \emph{Nature} \textbf{2020}, \emph{580},
  472--477\relax
\mciteBstWouldAddEndPuncttrue
\mciteSetBstMidEndSepPunct{\mcitedefaultmidpunct}
{\mcitedefaultendpunct}{\mcitedefaultseppunct}\relax
\EndOfBibitem
\bibitem[Sun \latin{et~al.}(2022)Sun, Ciarrocchi, Tagarelli, Gonzalez~Marin,
  Watanabe, Taniguchi, and Kis]{sun2022excitonic}
Sun,~Z.; Ciarrocchi,~A.; Tagarelli,~F.; Gonzalez~Marin,~J.~F.; Watanabe,~K.;
  Taniguchi,~T.; Kis,~A. Excitonic transport driven by repulsive dipolar
  interaction in a van der Waals heterostructure. \emph{Nat. Photonics}
  \textbf{2022}, \emph{16}, 79--85\relax
\mciteBstWouldAddEndPuncttrue
\mciteSetBstMidEndSepPunct{\mcitedefaultmidpunct}
{\mcitedefaultendpunct}{\mcitedefaultseppunct}\relax
\EndOfBibitem
\bibitem[Alexeev \latin{et~al.}(2019)Alexeev, Ruiz-Tijerina, Danovich, Hamer,
  Terry, Nayak, Ahn, Pak, Lee, Sohn, \latin{et~al.}
  others]{alexeev2019resonantly}
Alexeev,~E.~M.; Ruiz-Tijerina,~D.~A.; Danovich,~M.; Hamer,~M.~J.; Terry,~D.~J.;
  Nayak,~P.~K.; Ahn,~S.; Pak,~S.; Lee,~J.; Sohn,~J.~I., \latin{et~al.}
  Resonantly hybridized excitons in moir{\'e} superlattices in van der Waals
  heterostructures. \emph{Nature} \textbf{2019}, \emph{567}, 81--86\relax
\mciteBstWouldAddEndPuncttrue
\mciteSetBstMidEndSepPunct{\mcitedefaultmidpunct}
{\mcitedefaultendpunct}{\mcitedefaultseppunct}\relax
\EndOfBibitem
\bibitem[Hsu \latin{et~al.}(2019)Hsu, Lin, Lu, Lee, Chu, Li, Yao, Chang, and
  Shih]{hsu2019tailoring}
Hsu,~W.-T.; Lin,~B.-H.; Lu,~L.-S.; Lee,~M.-H.; Chu,~M.-W.; Li,~L.-J.; Yao,~W.;
  Chang,~W.-H.; Shih,~C.-K. Tailoring excitonic states of van der Waals
  bilayers through stacking configuration, band alignment, and valley spin.
  \emph{Sci. Adv.} \textbf{2019}, \emph{5}, eaax7407\relax
\mciteBstWouldAddEndPuncttrue
\mciteSetBstMidEndSepPunct{\mcitedefaultmidpunct}
{\mcitedefaultendpunct}{\mcitedefaultseppunct}\relax
\EndOfBibitem
\bibitem[Zhang \latin{et~al.}(2020)Zhang, Zhang, Wu, Wang, Gogna, Hou,
  Watanabe, Taniguchi, Kulkarni, Kuo, Forrest, and Deng]{zhang2020twist}
Zhang,~L.; Zhang,~Z.; Wu,~F.; Wang,~D.; Gogna,~R.; Hou,~S.; Watanabe,~K.;
  Taniguchi,~T.; Kulkarni,~K.; Kuo,~T.; Forrest,~S.~R.; Deng,~H. Twist-angle
  dependence of moir{\'e} excitons in ${\mathrm{WS}}_{2}$/${\mathrm{MoSe}}_{2}$
  heterobilayers. \emph{Nat. Commun.} \textbf{2020}, \emph{11}, 5888\relax
\mciteBstWouldAddEndPuncttrue
\mciteSetBstMidEndSepPunct{\mcitedefaultmidpunct}
{\mcitedefaultendpunct}{\mcitedefaultseppunct}\relax
\EndOfBibitem
\bibitem[Tang \latin{et~al.}(2021)Tang, Gu, Liu, Watanabe, Taniguchi, Hone,
  Mak, and Shan]{tang2021tuning}
Tang,~Y.; Gu,~J.; Liu,~S.; Watanabe,~K.; Taniguchi,~T.; Hone,~J.; Mak,~K.~F.;
  Shan,~J. Tuning layer-hybridized moir{\'e} excitons by the quantum-confined
  Stark effect. \emph{Nat. Nanotechnol.} \textbf{2021}, \emph{16}, 52--57\relax
\mciteBstWouldAddEndPuncttrue
\mciteSetBstMidEndSepPunct{\mcitedefaultmidpunct}
{\mcitedefaultendpunct}{\mcitedefaultseppunct}\relax
\EndOfBibitem
\bibitem[Yu and Yao(2021)Yu, and Yao]{yu2021luminescence}
Yu,~H.; Yao,~W. Luminescence Anomaly of Dipolar Valley Excitons in Homobilayer
  Semiconductor Moir\'e Superlattices. \emph{Phys. Rev. X} \textbf{2021},
  \emph{11}, 021042\relax
\mciteBstWouldAddEndPuncttrue
\mciteSetBstMidEndSepPunct{\mcitedefaultmidpunct}
{\mcitedefaultendpunct}{\mcitedefaultseppunct}\relax
\EndOfBibitem
\bibitem[MacNeill \latin{et~al.}(2015)MacNeill, Heikes, Mak, Anderson,
  Korm\'anyos, Z\'olyomi, Park, and Ralph]{macneill2015breaking}
MacNeill,~D.; Heikes,~C.; Mak,~K.~F.; Anderson,~Z.; Korm\'anyos,~A.;
  Z\'olyomi,~V.; Park,~J.; Ralph,~D.~C. Breaking of Valley Degeneracy by
  Magnetic Field in Monolayer ${\mathrm{MoSe}}_{2}$. \emph{Phys. Rev. Lett.}
  \textbf{2015}, \emph{114}, 037401\relax
\mciteBstWouldAddEndPuncttrue
\mciteSetBstMidEndSepPunct{\mcitedefaultmidpunct}
{\mcitedefaultendpunct}{\mcitedefaultseppunct}\relax
\EndOfBibitem
\bibitem[Li \latin{et~al.}(2014)Li, Ludwig, Low, Chernikov, Cui, Arefe, Kim,
  van~der Zande, Rigosi, Hill, Kim, Hone, Li, Smirnov, and Heinz]{li2014valley}
Li,~Y.; Ludwig,~J.; Low,~T.; Chernikov,~A.; Cui,~X.; Arefe,~G.; Kim,~Y.~D.;
  van~der Zande,~A.~M.; Rigosi,~A.; Hill,~H.~M.; Kim,~S.~H.; Hone,~J.; Li,~Z.;
  Smirnov,~D.; Heinz,~T.~F. Valley Splitting and Polarization by the Zeeman
  Effect in Monolayer ${\mathrm{MoSe}}_{2}$. \emph{Phys. Rev. Lett.}
  \textbf{2014}, \emph{113}, 266804\relax
\mciteBstWouldAddEndPuncttrue
\mciteSetBstMidEndSepPunct{\mcitedefaultmidpunct}
{\mcitedefaultendpunct}{\mcitedefaultseppunct}\relax
\EndOfBibitem
\bibitem[Aivazian \latin{et~al.}(2015)Aivazian, Gong, Jones, Chu, Yan, Mandrus,
  Zhang, Cobden, Yao, and Xu]{aivazian2015magnetic}
Aivazian,~G.; Gong,~Z.; Jones,~A.~M.; Chu,~R.-L.; Yan,~J.; Mandrus,~D.~G.;
  Zhang,~C.; Cobden,~D.; Yao,~W.; Xu,~X. Magnetic control of valley pseudospin
  in monolayer WSe2. \emph{Nat. Phys.} \textbf{2015}, \emph{11}, 148--152\relax
\mciteBstWouldAddEndPuncttrue
\mciteSetBstMidEndSepPunct{\mcitedefaultmidpunct}
{\mcitedefaultendpunct}{\mcitedefaultseppunct}\relax
\EndOfBibitem
\bibitem[Srivastava \latin{et~al.}(2015)Srivastava, Sidler, Allain, Lembke,
  Kis, and Imamo{\u{g}}lu]{srivastava2015valley}
Srivastava,~A.; Sidler,~M.; Allain,~A.~V.; Lembke,~D.~S.; Kis,~A.;
  Imamo{\u{g}}lu,~A. Valley Zeeman effect in elementary optical excitations of
  monolayer WSe2. \emph{Nat. Phys.} \textbf{2015}, \emph{11}, 141--147\relax
\mciteBstWouldAddEndPuncttrue
\mciteSetBstMidEndSepPunct{\mcitedefaultmidpunct}
{\mcitedefaultendpunct}{\mcitedefaultseppunct}\relax
\EndOfBibitem
\bibitem[Baek \latin{et~al.}(2020)Baek, Brotons-Gisbert, Koong, Campbell,
  Rambach, Watanabe, Taniguchi, and Gerardot]{baek2020highly}
Baek,~H.; Brotons-Gisbert,~M.; Koong,~Z.~X.; Campbell,~A.; Rambach,~M.;
  Watanabe,~K.; Taniguchi,~T.; Gerardot,~B.~D. Highly energy-tunable quantum
  light from moir\'e-trapped excitons. \emph{Sci. Adv.} \textbf{2020},
  \emph{6}, eaba8526\relax
\mciteBstWouldAddEndPuncttrue
\mciteSetBstMidEndSepPunct{\mcitedefaultmidpunct}
{\mcitedefaultendpunct}{\mcitedefaultseppunct}\relax
\EndOfBibitem
\bibitem[Wang \latin{et~al.}(2017)Wang, Shan, and Mak]{wang2017valley}
Wang,~Z.; Shan,~J.; Mak,~K.~F. Valley- and spin-polarized Landau levels in
  monolayer $\mathrm{WSe}_2$. \emph{Nat. Nanotechnol.} \textbf{2017},
  \emph{12}, 144--149\relax
\mciteBstWouldAddEndPuncttrue
\mciteSetBstMidEndSepPunct{\mcitedefaultmidpunct}
{\mcitedefaultendpunct}{\mcitedefaultseppunct}\relax
\EndOfBibitem
\bibitem[Liu \latin{et~al.}(2020)Liu, van Baren, Taniguchi, Watanabe, Chang,
  and Lui]{liu2020landau}
Liu,~E.; van Baren,~J.; Taniguchi,~T.; Watanabe,~K.; Chang,~Y.-C.; Lui,~C.~H.
  Landau-Quantized Excitonic Absorption and Luminescence in a Monolayer Valley
  Semiconductor. \emph{Phys. Rev. Lett.} \textbf{2020}, \emph{124},
  097401\relax
\mciteBstWouldAddEndPuncttrue
\mciteSetBstMidEndSepPunct{\mcitedefaultmidpunct}
{\mcitedefaultendpunct}{\mcitedefaultseppunct}\relax
\EndOfBibitem
\bibitem[Li \latin{et~al.}(2020)Li, Wang, Miao, Li, Lu, Jin, Lian, Meng, Blei,
  Taniguchi, Watanabe, Tongay, Yao, Smirnov, Zhang, and Shi]{li2020phonon}
Li,~Z. \latin{et~al.}  Phonon-exciton Interactions in ${\mathrm{WSe}}_{2}$
  under a quantizing magnetic field. \emph{Nat. Commun.} \textbf{2020},
  \emph{11}, 3104\relax
\mciteBstWouldAddEndPuncttrue
\mciteSetBstMidEndSepPunct{\mcitedefaultmidpunct}
{\mcitedefaultendpunct}{\mcitedefaultseppunct}\relax
\EndOfBibitem
\bibitem[Wang \latin{et~al.}(2020)Wang, Li, Lu, Li, Miao, Lian, Meng, Blei,
  Taniguchi, Watanabe, Tongay, Yao, Smirnov, Zhang, and
  Shi]{wang2020observation}
Wang,~T.; Li,~Z.; Lu,~Z.; Li,~Y.; Miao,~S.; Lian,~Z.; Meng,~Y.; Blei,~M.;
  Taniguchi,~T.; Watanabe,~K.; Tongay,~S.; Yao,~W.; Smirnov,~D.; Zhang,~C.;
  Shi,~S.-F. Observation of Quantized Exciton Energies in Monolayer
  ${\mathrm{WSe}}_{2}$ under a Strong Magnetic Field. \emph{Phys. Rev. X}
  \textbf{2020}, \emph{10}, 021024\relax
\mciteBstWouldAddEndPuncttrue
\mciteSetBstMidEndSepPunct{\mcitedefaultmidpunct}
{\mcitedefaultendpunct}{\mcitedefaultseppunct}\relax
\EndOfBibitem
\bibitem[Li \latin{et~al.}(2020)Li, Goryca, Wilson, Stier, Xu, and
  Crooker]{li2020spontaneous}
Li,~J.; Goryca,~M.; Wilson,~N.~P.; Stier,~A.~V.; Xu,~X.; Crooker,~S.~A.
  Spontaneous Valley Polarization of Interacting Carriers in a Monolayer
  Semiconductor. \emph{Phys. Rev. Lett.} \textbf{2020}, \emph{125},
  147602\relax
\mciteBstWouldAddEndPuncttrue
\mciteSetBstMidEndSepPunct{\mcitedefaultmidpunct}
{\mcitedefaultendpunct}{\mcitedefaultseppunct}\relax
\EndOfBibitem
\bibitem[Smole\ifmmode~\acute{n}\else \'{n}\fi{}ski
  \latin{et~al.}(2019)Smole\ifmmode~\acute{n}\else \'{n}\fi{}ski, Cotlet,
  Popert, Back, Shimazaki, Kn\"uppel, Dietler, Taniguchi, Watanabe, Kroner, and
  Imamoglu]{smolenski2019interaction}
Smole\ifmmode~\acute{n}\else \'{n}\fi{}ski,~T.; Cotlet,~O.; Popert,~A.;
  Back,~P.; Shimazaki,~Y.; Kn\"uppel,~P.; Dietler,~N.; Taniguchi,~T.;
  Watanabe,~K.; Kroner,~M.; Imamoglu,~A. Interaction-Induced Shubnikov--de Haas
  Oscillations in Optical Conductivity of Monolayer ${\mathrm{MoSe}}_{2}$.
  \emph{Phys. Rev. Lett.} \textbf{2019}, \emph{123}, 097403\relax
\mciteBstWouldAddEndPuncttrue
\mciteSetBstMidEndSepPunct{\mcitedefaultmidpunct}
{\mcitedefaultendpunct}{\mcitedefaultseppunct}\relax
\EndOfBibitem
\bibitem[Klein \latin{et~al.}(2021)Klein, H\"otger, Florian, Steinhoff,
  Delhomme, Taniguchi, Watanabe, Jahnke, Holleitner, Potemski, Faugeras,
  Finley, and Stier]{klein2021controlling}
Klein,~J.; H\"otger,~A.; Florian,~M.; Steinhoff,~A.; Delhomme,~A.;
  Taniguchi,~T.; Watanabe,~K.; Jahnke,~F.; Holleitner,~A.~W.; Potemski,~M.;
  Faugeras,~C.; Finley,~J.~J.; Stier,~A.~V. Controlling exciton many-body
  states by the electric-field effect in monolayer ${\mathrm{MoS}}_{2}$.
  \emph{Phys. Rev. Research} \textbf{2021}, \emph{3}, L022009\relax
\mciteBstWouldAddEndPuncttrue
\mciteSetBstMidEndSepPunct{\mcitedefaultmidpunct}
{\mcitedefaultendpunct}{\mcitedefaultseppunct}\relax
\EndOfBibitem
\bibitem[Thomas and Hopfield(1961)Thomas, and Hopfield]{thomas1961amagneto}
Thomas,~D.~G.; Hopfield,~J.~J. A Magneto-Stark Effect and Exciton Motion in
  CdS. \emph{Phys. Rev.} \textbf{1961}, \emph{124}, 657--665\relax
\mciteBstWouldAddEndPuncttrue
\mciteSetBstMidEndSepPunct{\mcitedefaultmidpunct}
{\mcitedefaultendpunct}{\mcitedefaultseppunct}\relax
\EndOfBibitem
\bibitem[Hopfield and Thomas(1961)Hopfield, and Thomas]{hopfield1961fine}
Hopfield,~J.~J.; Thomas,~D.~G. Fine Structure and Magneto-Optic Effects in the
  Exciton Spectrum of Cadmium Sulfide. \emph{Phys. Rev.} \textbf{1961},
  \emph{122}, 35--52\relax
\mciteBstWouldAddEndPuncttrue
\mciteSetBstMidEndSepPunct{\mcitedefaultmidpunct}
{\mcitedefaultendpunct}{\mcitedefaultseppunct}\relax
\EndOfBibitem
\bibitem[Gorkov and Dzyaloshinskii(1968)Gorkov, and
  Dzyaloshinskii]{gorkov1968contribution}
Gorkov,~L.~P.; Dzyaloshinskii,~I.~E. Contribution to the Theory of the Mott
  Exciton in a Strong Magnetic Field. \emph{Sov. Phys. JETP} \textbf{1968},
  \emph{26}, 449--451\relax
\mciteBstWouldAddEndPuncttrue
\mciteSetBstMidEndSepPunct{\mcitedefaultmidpunct}
{\mcitedefaultendpunct}{\mcitedefaultseppunct}\relax
\EndOfBibitem
\bibitem[Chang \latin{et~al.}(2002)Chang, Xia, Wu, Feng, and
  Peeters]{chang2002quantum}
Chang,~K.; Xia,~J.~B.; Wu,~H.~B.; Feng,~S.~L.; Peeters,~F.~M. Quantum-confined
  magneto-Stark effect in diluted magnetic semiconductor coupled quantum wells.
  \emph{Appl. Phys. Lett.} \textbf{2002}, \emph{80}, 1788--1790\relax
\mciteBstWouldAddEndPuncttrue
\mciteSetBstMidEndSepPunct{\mcitedefaultmidpunct}
{\mcitedefaultendpunct}{\mcitedefaultseppunct}\relax
\EndOfBibitem
\bibitem[Karin \latin{et~al.}(2016)Karin, Linpeng, Glazov, Durnev, Ivchenko,
  Harvey, Rai, Ludwig, Wieck, and Fu]{karin2016giant}
Karin,~T.; Linpeng,~X.; Glazov,~M.~M.; Durnev,~M.~V.; Ivchenko,~E.~L.;
  Harvey,~S.; Rai,~A.~K.; Ludwig,~A.; Wieck,~A.~D.; Fu,~K.-M.~C. Giant
  permanent dipole moment of two-dimensional excitons bound to a single
  stacking fault. \emph{Phys. Rev. B} \textbf{2016}, \emph{94}, 041201\relax
\mciteBstWouldAddEndPuncttrue
\mciteSetBstMidEndSepPunct{\mcitedefaultmidpunct}
{\mcitedefaultendpunct}{\mcitedefaultseppunct}\relax
\EndOfBibitem
\bibitem[Farenbruch \latin{et~al.}(2020)Farenbruch, Mund, Fr\"ohlich, Yakovlev,
  Bayer, Semina, and Glazov]{farenbruch2020magneto}
Farenbruch,~A.; Mund,~J.; Fr\"ohlich,~D.; Yakovlev,~D.~R.; Bayer,~M.;
  Semina,~M.~A.; Glazov,~M.~M. Magneto-Stark and Zeeman effect as origin of
  second harmonic generation of excitons in ${\mathrm{Cu}}_{2}\mathrm{O}$.
  \emph{Phys. Rev. B} \textbf{2020}, \emph{101}, 115201\relax
\mciteBstWouldAddEndPuncttrue
\mciteSetBstMidEndSepPunct{\mcitedefaultmidpunct}
{\mcitedefaultendpunct}{\mcitedefaultseppunct}\relax
\EndOfBibitem
\bibitem[Lafrentz \latin{et~al.}(2013)Lafrentz, Brunne, Kaminski, Pavlov,
  Rodina, Pisarev, Yakovlev, Bakin, and Bayer]{lafrentz2013magneto}
Lafrentz,~M.; Brunne,~D.; Kaminski,~B.; Pavlov,~V.~V.; Rodina,~A.~V.;
  Pisarev,~R.~V.; Yakovlev,~D.~R.; Bakin,~A.; Bayer,~M. Magneto-Stark Effect of
  Excitons as the Origin of Second Harmonic Generation in ZnO. \emph{Phys. Rev.
  Lett.} \textbf{2013}, \emph{110}, 116402\relax
\mciteBstWouldAddEndPuncttrue
\mciteSetBstMidEndSepPunct{\mcitedefaultmidpunct}
{\mcitedefaultendpunct}{\mcitedefaultseppunct}\relax
\EndOfBibitem
\bibitem[Herold \latin{et~al.}(1981)Herold, Ruder, and Wunner]{herold1981the}
Herold,~H.; Ruder,~H.; Wunner,~G. The two-body problem in the presence of a
  homogeneous magnetic field. \emph{J. Phys. B: At. Mol. Phys.} \textbf{1981},
  \emph{14}, 751--764\relax
\mciteBstWouldAddEndPuncttrue
\mciteSetBstMidEndSepPunct{\mcitedefaultmidpunct}
{\mcitedefaultendpunct}{\mcitedefaultseppunct}\relax
\EndOfBibitem
\bibitem[Amorim \latin{et~al.}(2016)Amorim, Cortijo, {de Juan}, Grushin,
  Guinea, Gutiérrez-Rubio, Ochoa, Parente, Roldán, San-Jose, Schiefele,
  Sturla, and Vozmediano]{amorim2016novel}
Amorim,~B.; Cortijo,~A.; {de Juan},~F.; Grushin,~A.; Guinea,~F.;
  Gutiérrez-Rubio,~A.; Ochoa,~H.; Parente,~V.; Roldán,~R.; San-Jose,~P.;
  Schiefele,~J.; Sturla,~M.; Vozmediano,~M. Novel effects of strains in
  graphene and other two dimensional materials. \emph{Phys. Rep.}
  \textbf{2016}, \emph{617}, 1--54\relax
\mciteBstWouldAddEndPuncttrue
\mciteSetBstMidEndSepPunct{\mcitedefaultmidpunct}
{\mcitedefaultendpunct}{\mcitedefaultseppunct}\relax
\EndOfBibitem
\bibitem[Kang \latin{et~al.}(2013)Kang, Tongay, Zhou, Li, and Wu]{kang2013band}
Kang,~J.; Tongay,~S.; Zhou,~J.; Li,~J.; Wu,~J. Band offsets and
  heterostructures of two-dimensional semiconductors. \emph{Appl. Phys. Lett.}
  \textbf{2013}, \emph{102}, 012111\relax
\mciteBstWouldAddEndPuncttrue
\mciteSetBstMidEndSepPunct{\mcitedefaultmidpunct}
{\mcitedefaultendpunct}{\mcitedefaultseppunct}\relax
\EndOfBibitem
\bibitem[Fang \latin{et~al.}(2018)Fang, Carr, Cazalilla, and
  Kaxiras]{fang2018electronic}
Fang,~S.; Carr,~S.; Cazalilla,~M.~A.; Kaxiras,~E. Electronic structure theory
  of strained two-dimensional materials with hexagonal symmetry. \emph{Phys.
  Rev. B} \textbf{2018}, \emph{98}, 075106\relax
\mciteBstWouldAddEndPuncttrue
\mciteSetBstMidEndSepPunct{\mcitedefaultmidpunct}
{\mcitedefaultendpunct}{\mcitedefaultseppunct}\relax
\EndOfBibitem
\end{mcitethebibliography}

\end{document}